\mag=\magstephalf
\pageno=1
\input amstex
\documentstyle{amsppt}
\TagsOnRight
\interlinepenalty=1000
\hsize=6.6truein
\voffset=23pt
\vsize =9.8truein
\advance\vsize by -\voffset
\nologo

\def\tvskip{\vskip 0.5 cm}

\font\twobf=cmbx12
      
\define \RR{\Bbb R}

\define \SS{\Cal S}

\define \HK{{\text{HK}}}
\define \Dirac{{\text{Dirac}}}

\define \laDD{\overset{\leftarrow}\to{/\!\!\!\!\Cal{D}}}

\define \DD{/\!\!\!\!\Cal{D}}

\define \CCD{\Cal{D}}

\define \ii{\roman i}
\define \ee{\roman e}
\define \dd{\roman d}

\def\mnarrower{\advance\leftskip by 50pt 
\advance\rightskip by 50pt}

{\centerline{\bf{Immersion Anomaly of Dirac Operator on 
Surface in $\Bbb R^3$}}}
\author
\endauthor
\affil
Shigeki MATSUTANI\\
2-4-11 Sairenji, Niihama, Ehime, 792 Japan \\
\endaffil
\endtopmatter


\tvskip
\centerline{\twobf Abstract}
\tvskip

In previous report (J. Phys. A (1997) {\bf 30} 4019-4029), 
I showed that the Dirac field  confined in a surface immersed 
in $\Bbb R^3$
by means of a mass type potential is governed by the 
Konopelchenko-Kenmotsu-Weierstrass-Enneper equation.
In this article, I quantized the Dirac field and calculated 
the gauge transformation which exhibits the gauge freedom of 
the parameterization
of the surface.
Then using the Ward-Takahashi identity, I showed that 
the expectation value of the action of the Dirac field is 
expressed by the Willmore functional and area of the surface.


\document

\tvskip
\centerline{\twobf \S 1. Introduction}
\tvskip

In the previous report [1], I showed that the Dirac field 
confined in 
a thin curved surface $S$ immersed in three dimensional flat 
space
$\Bbb R^3$ obeys the Dirac 
equation which is discovered by Konopelchenko [2-4]
$$
    \partial f_1 = p f_2, \quad \bar \partial f_2= -pf_1,
        \tag 1-1
$$
where
$$
      p:= \frac{1}{2}\sqrt{\rho} H , \tag 1-2
$$
$H$ is the mean curvature of the surface $S$ parameterized by 
complex $z$ 
and $\rho$ is the factor of the conformal metric induced from 
$\Bbb R^3$.
 
This equation completely represents the immersed geometry as 
the old Weierstrass-Enneper  equation expresses the minimal 
surface [2].

Even though the relation had been essentially found by Kenmotsu
 [4-7],
the formulation as the Dirac type was performed by Konopelchenko
and recently it is revealed that the Dirac operator  
has more physical and mathematical meanings;
the Dirac operator is a translator between the geometrical and
analytical objects [8] even in the arithmetic geometry of the 
number theory [9].
Thus although the Dirac type equation (1-1) has been  called as 
the generalized Weierstrass equation, in this article
I will call it Konopelchenko-Kenmotsu-Weierstrass-Enneper (KKWE) 
equation.

The immersion geometry is currently studied in the various fields,
{\it e.g.}, soliton theory, the differential geometry, the
harmonic map theory, string theory and so on.
In the soliton theory, the question what is the integrable
is the most important theme and one of its answers might be 
found in the immersed geometry. In fact, the static sine-Gordon
equation was discovered by Euler, from the energy functional
of the elastica given by Daniel Bernoulli in eighteenth century,
 as
an elastica immersed in $\Bbb R^2$ [10] and the net sine-Gordon
equation was found in the last century as a surface immersed
in $\Bbb R^3$ [11]. Recently Goldstein and Petrich
discovered the modified KdV (MKdV) hierarchy by considering
one parameter deformation of a space curve immersed in $\Bbb R^2$
 [12,13].
After their new interpretation, there appear several
geometrical realizations of the soliton theory [14-16].
In the differential geometry,
after the discovery the exotic solution of the constant 
mean curvature surface by Wente [17], the extrinsic structure is
currently studied again [18-20].
In the harmonic map theory, it is found that the minimal point of
a functional is, sometimes, integrable and  classified by the
extrinsic topology  [20] like the elastica
problem [21], a prototype of such model  which firstly entered 
in history [10].

Furthermore Polyakov introduced an extrinsic action in the string
theory and the theory of 2-dimensional gravity for 
renomalizability [22].
His program has been studied in the framework of W-algebra [23] 
but recently  was investigated by Carroll and Konopelchenko [24]
and Viswanathan and Parthasarathy [25] using more direct method.
Polyakov's extrinsic action in the classical level is the same as
the Willmore functional [26,27],
$$
      W = \int_{\Cal S}  \dd \text{vol}\  H^2 , \tag 1-3
$$
where "$\dd \text{vol}$" is a volume form of the surface $S$.

Accordingly the immersed surface is a very current object
and its studies currently progress.

On the other hand
I have been studying the Dirac field confined in an immersed 
object
and its relationship with the immersed object itself [28-32].
Since the Dirac operator should be regarded as a translator 
(a functor) between the analytical object and the geometrical 
object
 [8-9], in terms of the Dirac operator,
I have been studying the physical and geometrical meanings of
the abstract theorems in the soliton theory and quantum theory
focusing on the elastica problem [28-32] and recently on the 
immersed surface [1].

In this article, I will deal with the quantized Dirac field and 
investigate the gauge freedom which does not change the Willmore
functional. In other words, I will search for a symmetry in the 
classical
level and compute its anomalous relation in  quantum level
using  Fujikawa type prescription [33-35]. Furthermore since 
there are negative
eigenvalues which makes the theory worse in a calculus,
I will propose new regularization which can be regarded as a 
local version
of a  generalization of Hurwitz $\zeta$-function [36] and
then I will obtain the finite result and 
the coupling constant of the Liouville action [22] as 
convergent parameter.
Finally I have the relation between the expectation value of the 
action of the Dirac field and the Willmore functional.
It reminds me of the boson-fermion correspondence in this system.
It should be noted that even though it does not directly generate
topological index,
it could be regarded as a local version of the Atiyah-Singer 
type index theorem [8,9,37-39].

Organization of this article is as follows.
Section 2 reviews the extrinsic geometry of a surface 
immersed in $\Bbb R^3$.
There I will introduce the Willmore surface as a free energy 
of a thin elastic surface. 
In section 3, starting with the quantized Dirac field whose 
on-shell equation is KKWE equation (1-1), I will calculate the 
variation of a
gauge transformation. Using the Ward-Takahashi identity,
I will obtain an anomalous relation exhibiting this system.
In section 4, I will discuss the obtained results.

\tvskip
\centerline{\twobf \S 2. Conformal Surface Immersed in $\RR^3$}
\tvskip

In this article, I will consider a compact surface $\SS$ 
immersed in $\Bbb R^3$,
which has a complex structure [1-4],
$$
    \varpi:\Sigma \to \SS \subset \Bbb R^3,     \tag 2-1
$$
where $\Sigma$ and $\SS$ are two-dimensional conformal manifold.
$\SS$ is parameterized by two-dimensional coordinate system 
$(q^1,q^2) \in \Sigma$.

A position on a conformal compact surface
$\SS$  is represented using the affine vector 
${\bold x}(q^1,q^2)$$=(x^I)$
$ =(x^1,x^2,x^3)$ in $\RR^3$
and the normal unit vector of $ \SS $  is denoted by 
${\bold e}_3 $.
I sometimes regard the euclidean space as a product manifold 
of complex plane and 
real line, $\Bbb R^3 \approx \Bbb C \times \Bbb R$ [1-4],
$$
    Z :=x^1 +\ii x^2\in \Bbb C, \quad x^3 \in \Bbb R .  \tag 2-2
$$
The surface $\SS$ has  the conformal flat metric,
$$
    g_{\alpha \beta} \dd q^\alpha \dd q^\beta=\rho 
    \delta_{\alpha, \beta} 
    \dd q^\alpha \dd q^\beta . \tag 2-3
$$
The complex parameterization of the surface is employed,
$$
    z:=q^1 +\ii q^2 , \tag 2-4
$$
and
$$
    \partial := \frac{1}{2}(\partial_{q^1}-\ii \partial_{q^2}), 
    \qquad
    \bar \partial:= \frac{1}{2}(\partial_{q^1}+
    \ii \partial_{q^2}) ,
    \quad \dd^2 q:=\dd q^1\dd q^2
    =: \frac{1}{2} \dd^2 z:=\frac{1}{2} \ii\dd z \dd \bar z. 
     \tag 2-5
$$
For a given function $f$ over $\SS$, if $f$ is real analytic, 
I denote it as $f=f(q)$ but if it should be regarded as a 
complex analytic function,
I will use the notation, $f=f(z)$.

Then  the moving frame can be written as 
$$
     e^I_{\ \alpha}:= \partial_\alpha x^I,\quad
     e^I_{\ z}:=\partial x^I,  \tag 2-6
$$
where $\partial_\alpha := \partial_{q^\alpha}:=\partial
/\partial q^\alpha$.
Their inverse matrices are denoted as $e^\alpha_{\ I}$ and 
$e^z_{\ I}$. The metric is expressed as
$$
  \frac{1}{4}  \rho = <\bold e_z, \bold e_{\bar z}>:= 
  \delta_{a,b} e^I_{\ z} e^b_{\ \bar z}. \tag 2-7
$$
 Here $<,>$ denotes the canonical inner product in the 
 euclidean space $\RR^3$.

The second fundamental form is denoted as,
$$
          \gamma^3_{\ \beta\alpha}
        :=<\!{\bold e}_3,\partial_\alpha {\bold e}_\beta\!>.  
        tag 2-8
$$
Using the relation 
$<\!{\bold e}_3,\partial_\alpha{\bold e}_3\!>=0$,
the Weingarten map, 
$-\gamma^\alpha_{\ \beta 3}{\bold e}_\alpha$, is defined by 
$$
    \gamma^\alpha_{\ \beta 3}
     =<\!{\bold e}^\alpha,\partial_\beta{\bold e}_3\!> . 
     \tag 2-9
$$
From  $\partial_\alpha\!\!<\!{\bold e}^\gamma,{\bold e}_3\!>=0$,
 $\gamma^\alpha_{\ \beta 3}$ is related to 
 the second fundamental form through the relation,
$$
        \gamma^3_{\ \beta\alpha} = 
            -\gamma^\gamma_{\ 3\alpha}g_{\gamma\beta} 
          = -\gamma^\beta_{\ 3\alpha} \rho
                           ,  \tag 2-10
$$
is the surface metric. 
It is worth while noting that for a scaling of 
$(q^1,q^2)\to \lambda(q^1,q^2)$,
the Weingarten map does not change.

In terms of the Weingarten map, I will introduce invariant 
quantities for the coordinate transformation if I fix the 
surface $\SS$. They are known as 
 the mean and the Gaussian curvatures on $\SS$:
$$
H:=-\frac{1}{2}\roman{tr}_2(\gamma^\alpha_{\ 3\beta}), \qquad
    K:= \roman{det}_2(\gamma^\alpha_{\ 3\beta}).    
     \tag 2-11
$$
Here  $\roman{tr}_2$ and $\det_2$
are the two-dimensional trace and determinant over $\alpha$ 
and $\beta$ respectively. 
Due to the Gauss's egregium theorem, I have the relation,
$$
        K =-2 \frac{1}{\rho}\partial \bar \partial \log \rho,
         \tag 2-12
$$
and from the properties of complex manifold, I obtain
$$
    H=\frac{2}{\rho} <\partial \bar \partial \bold x, \bold e^3>
              =\frac{4}{\ii \rho^2}\epsilon_{IJK}
          \partial \bar \partial x^I\partial x^J 
          \bar \partial x^K.
                \tag 2-13
$$

Using the independence of the choice of the local coordinate,
 I will introduce a
proper coordinate transformation which diagonalizes the 
Weingarten map,
$$
      U_{\alpha}^{T\ \gamma} \gamma^\alpha_{\ 3\beta}
       U_{\delta}^{\ \beta}=\text{diag}(k_1,k_2). \tag 2-14
$$
These diagonal elements $(k_1,k_2)$ 
are known as the principal curvatures of the surface $\SS$.
In terms of these values, the Gauss and mean curvatures are 
expressed as [27],
$$
      K = k_1 k_2 , \quad H=\frac{1}{2}(k_1+k_2). \tag 2-15
$$

Here I will regarded the surface $\SS$ as a shape of a thin 
elastic surface. Its local free energy density is given 
as an invariant for the local coordinate transformation. 
On the other hand, the difference of the local surface 
densities between inside and outside
surfaces is proportional to the extrinsic curvature due to 
its thickness,
for a local deformation of the surface. By the linear 
response of the elastic body theory and independence of 
the coordinate transformation, 
the free energy might be given as [40]
$$
      f=B_0 H^2 + B_1 K=\frac{1}{4}B_0(k_1^2 +k_2^2)
      +(B_0+B_1)k_1 k_2 , \tag 2-16
$$
where $B$'s are elastic constants.
However using the Gauss-Bonnet theorem [27], the integral of 
the second term in (2-16) is expressed as,
$$
      \int \rho \dd^2 q K=\int \rho \dd^2 q k_1 k_2
      = 2\pi \chi,  \tag 2-17
$$
where $\chi$ is the Euler characteristic,  which is an integer 
and exhibits the global topological properties of the surface.
 Hence the second term in (2-16)
is not dynamical one if one fixes the topology of the system.

Hence the free energy of the system  becomes 
$$
      W = B_0 \int \rho \dd^2 q\  H^2. \tag 2-18
$$
This functional integral is known as the Willmore functional 
[26,27] and,
recently, as the Polyakov's extrinsic action in the 
2-dimensional gravity [22,24,25].
For later convenience, I will fix $B_0=1$ and introduce a 
quantity [1-4],
$$
      p:= \frac{1}{2}\sqrt{\rho} H=\frac{1}{2}g^{1/4} H. 
      \tag 2-19
$$
Using this new quantities, the Willmore functional is written as,
$$
      W= 4 \int \dd^2 q\ p^2. \tag 2-20
$$

\tvskip
\centerline{\twobf \S 3. Immersion Anomaly}
\tvskip

In the previous report [1], I showed that the Dirac field 
confined in the surface $\SS$
obeys the KKWE equation (1-1).
In this article, I will deal with the quantized fermion over 
the immersed surfaces.
As I did in ref.[29], after I quantize the Dirac field,
I can obtain the quantized Dirac field over the immersed thin
surface $\SS$ by confining it using the confinement mass-type 
potential.
This computation can be performed parallelled to the arguments 
in refs. [1] and [29].
Thus in this article, I will start with the quantized
Dirac field of the surface $\SS$ [1,22,41]. The partition 
function of the Dirac field is given as,
$$
        Z[\bar{\pmb{ \psi}},\pmb{ \psi},\rho,H]
         = \int D  \bar{\pmb{ \psi}}  D\pmb{ \psi} 
         \exp\left(-S_\Dirac[ \bar{\pmb{ \psi}} ,
         \pmb{ \psi},\rho,H]\right) ,\tag 3-1
$$
where [1,15,42]
$$
     S_\Dirac[ \bar{\pmb{ \psi}} ,\pmb{ \psi},\rho,H] 
     = \int \rho \dd^2 q \ 
     \Cal L_\Dirac[ \bar{\pmb{ \psi}} ,\pmb{ \psi},\rho,H],
     \quad
     \Cal L_\Dirac[ \bar{\pmb{ \psi}} ,\pmb{ \psi},\rho,H]
      =\ii \bar{\pmb{ \psi}} \DD \pmb{ \psi}, \tag 3-2
$$
$$
 \DD:=\gamma^\alpha \CCD_\alpha + \gamma^3 H, \quad
    \CCD_\alpha:=\partial_\alpha +\omega_\alpha, \quad 
     \omega_{\alpha}:=- \frac{1}{4} \rho^{-1}\sigma^{ab}
     (\partial_a  \rho \delta_{\alpha b}
    -\partial_b  \rho \delta_{\alpha a}) ,\tag 3-3
$$
$$
    \gamma^\alpha=e_a^{\ \alpha} \sigma^a,\quad  
    \gamma^3=\sigma^3,\quad
            \sigma^{ab}:=[\sigma^a,\sigma^b]/2, \tag 3-4
$$
$$
     \bar{\pmb{ \psi}}={\pmb{ \psi}}^\dagger \sigma^1 \rho^{1/2} 
     .\tag 3-5
$$
Here I denoted the Pauli matrix as $\sigma^a$ and used the 
conformal gauge freedom, 
$e_\alpha^a = \rho^{1/2} \delta_\alpha^a$. The indices 
$a,b$ is of the inner space and run over 1 and 2. 
The Dirac operator can be expressed as
$$
\split
     \DD&:=(\gamma^\alpha \CCD_\alpha + \gamma^3 H) \\
            &=\sigma^a\rho^{-1/2} \delta^\alpha_{\ a}[
        \partial_\alpha + \frac{1}{2}\rho^{-1} 
        (\partial_\alpha \rho)]
          +\sigma^3 H\\
           &=2\pmatrix H/2 & \rho^{-1/2} 
           \bar \partial \rho^{1/2} \\
                 \rho^{-1/2}  \partial \rho^{1/2} & -H/2 
               \endpmatrix     . \endsplit
        \tag 3-6
$$
Noting the fact that ${\pmb{ \psi}}$'s are just integral 
variables in the path integral,
the kinetic term of the Dirac operator is hermite,
$$
    \split
    <{\pmb{ \psi}}|\ii \pmatrix 0 & \rho^{-1} 
    \bar \partial \rho^{1/2} \\
                 \rho^{-1}  \partial \rho^{1/2} & 0 \endpmatrix 
         {\pmb{ \psi}}>&=
   \ii \int \rho \dd^2 z\ \bar{\pmb{ \psi}}
         \pmatrix 0 & \rho^{-1} \bar \partial \rho^{1/2} \\
        \rho^{-1}  \partial \rho^{1/2} & 0 \endpmatrix 
        {\pmb{ \psi}}\\
 &= -\ii \int \dd^2 z\ \left(  \bar{\pmb{ \psi}}_1 
        \overset{\leftarrow}\to{ \bar \partial} 
        \rho^{1/2}\pmb{ \psi}_2
        + \bar{\pmb{ \psi}}_2^* 
        \overset{\leftarrow}\to{ \partial} 
      \rho^{1/2}\pmb{ \psi}_1\right)\\
     &=-\ii \int \dd^2 z\ \left( {\pmb{ \psi}}_2^*\rho^{1/2} 
           \overset{\leftarrow}\to{\bar  \partial} \rho^{1/2}
           \pmb{ \psi}_2
           +  {\pmb{ \psi}}_1\rho^{1/2} \overset{\leftarrow}
           \to{ \partial} 
      \rho^{1/2}\pmb{ \psi}_1 \right)\\
      &=\int \rho \dd^2 z\ \left(\ii\pmatrix 0 & \rho^{-1}  
      \partial \rho^{1/2} \\
                 \rho^{-1} \bar \partial \rho^{1/2} & 0 
               \endpmatrix {\pmb{ \psi}}\right)^\dagger 
               \sigma^1 \rho^{1/2} {\pmb{ \psi}}\\
           &\sim  <\ii \pmatrix 0 & \rho^{-1} \bar 
           \partial \rho^{1/2} \\
                 \rho^{-1}  \partial \rho^{1/2} & 0 
               \endpmatrix{\pmb{ \psi}} |{\pmb{ \psi}}>.
             \endsplit \tag 3-7
$$

As I showed in ref.[1], when I redefine the Dirac field in 
the surface $\SS$ as
$$
    f:=\rho^{1/2} \pmb{ \psi}  ,\tag 3-8
$$
 the Dirac operator  becomes simpler,
$$
      \split
    \Cal L_f \sqrt{g} \dd^2 q 
    &=\ii f^\dagger \sigma^1(\sigma^a\delta^\alpha_{\ a}
    \partial_\alpha
    + \rho^{1/2} H \sigma^3)f\  \dd^2 q \\
    &= \ii \bar f\pmatrix p  &\bar \partial  \\
                   \partial & -p \endpmatrix f
    \ \dd^2 z, \endsplit \tag 3-9
$$
where $p$ is defined in (2-21)

Then the KKWE equation is obtained as the on-shell motion 
of (3-9) [1-4],
$$
    \partial f_1 = p f_2, \quad \bar \partial f_2= -pf_1.
        \tag 3-10
$$
These equations which were found by Konopelchenko reproduce 
all properties of the extrinsic geometry of this system. 
Their properties were studied
by Konopelchenko and Taimanov [2-4,6,7],
$$
      f_1 = \sqrt{\ii \bar \partial \bar Z/2}, \quad f_2= 
      \sqrt{-\ii  \partial \bar Z/2}. \tag 3-11
$$
This relation may be interpreted as the bosonaization in 
the conformal field theory [22].
It should be noted that its lower dimensional version of 
the KKWE equation
is found through study of the elastica as the square root of the 
Frenet-Serret relation [12-14].

The Willmore functional (2-20) is expressed by $p$ and
$p$ consists of multiple of $H$ and $\sqrt{\rho}$. Hence
by fixing $p$, there still remains a freedom of choice of
$\rho$; fixing $p$ means the deformation of $\rho$
without changing the Willmore functional.
Corresponding to the deformation preserving the
value of the Willmore functional, 
the lagrangian of the Dirac field (3-2) has a similar
gauge freedom which does not change the action $S_\Dirac$.
In fact using such the gauge freedom, I scaled the Dirac field 
$\pmb{\psi}$ to $f$ in (3-8).

However in the quantum field theory,
even though the lagrangian is invariant for a transformation,
the partition function is not in general due to the jacobian of 
the functional measure. The purpose of this article is 
to calculate this quantum effect.
Thus I will estimate the infinitesimal gauge transformation
which does not change the action of the Dirac field (3-2) and
is an analogue of the transformation of (3-8).

Following a conventional notation, I will introduce the 
dilatation parameter,
$$
      \phi:= \frac{1}{2}\log \rho, \tag 3-13
$$ 
which is sometimes called as dilaton [22]. 
Furthermore I will rewrite the Dirac operator in (3-6) as, 
$$
 \DD=2\pmatrix p\rho^{-1/2} & \rho^{-1} \bar \partial 
 \rho^{1/2} \\
           \rho^{-1}  \partial \rho^{1/2} & -p\rho^{-1/2} 
           \endpmatrix
                . \tag 3-12 
$$

As I mentioned above,
I will deal with the variation of the dilaton preserving $p$,
$$
       \quad \phi \to \phi + \alpha,\ \  
       (\rho \to \rho\ee^{2 \alpha}),\quad
      p \to p . \tag 3-14
$$
For the infinitesimal variation of the dilaton, 
the action of the fermionic field changes its value,
$$
      \Cal S_\Dirac \to {\Cal S_\Dirac}'=\Cal S_\Dirac +\ii
       \int \rho \dd^2 q \ \alpha(\rho^{-1} 
       \delta^{\beta}_a\partial_\beta \bar{\pmb{ \psi}}\sigma^a 
       \rho^{1/2}\pmb{ \psi} +
      \bar{\pmb{ \psi}}\DD\pmb{ \psi}  ) .
      \tag 3-15
$$
However this change can be classically canceled out by 
the gauge transformation,
$$
 {\pmb{ \psi}} \to  {\pmb{ \psi}}'= \ee^{-\alpha}  
 {\pmb{ \psi}}, \quad
 \bar {\pmb{ \psi}} \to  \bar {\pmb{ \psi}}^{\prime}= 
  \bar{\pmb{ \psi}}. \tag 3-16
$$
In other words, I have the identity,
$$
 S_\Dirac[ \bar{\pmb{ \psi}} ,\pmb{ \psi},\rho',p] \to
  S_\Dirac[ \bar{\pmb{ \psi}}' ,\pmb{ \psi}',\rho',p]
 =S_\Dirac[ \bar{\pmb{ \psi}} ,\pmb{ \psi},\rho,p]. \tag 3-17
$$

Here I will evaluate the variations (3-14) and (3-16) in the
 framework of  the quantum theory [30,31,33],
$$
    \split
    Z[\rho',H']& =\int\roman  D\bar{\pmb{ \psi}}
    \roman D{\pmb{ \psi}} 
        \exp(-S_\Dirac[ \bar{\pmb{ \psi}} ,
        \pmb{ \psi},\rho',p] )=
         :Z_1 \\ 
          &=\int  D\bar{\pmb{ \psi}}'\roman D{\pmb{ \psi}}' 
         \exp(-S_\Dirac[ \bar{\pmb{ \psi}}' ,
         \pmb{ \psi}',\rho',p] )\\
          &=\int  D\bar{\pmb{ \psi}}\roman D{\pmb{ \psi}}
        \frac{\delta \pmb{ \psi} \delta \bar{\pmb{ \psi}}}
        {\delta \pmb{ \psi}' \delta \bar{\pmb{ \psi}}'}
         \exp(-S_\Dirac[ \bar{\pmb{ \psi}} ,
         \pmb{ \psi},\rho,p] )=:Z_2 . \\ 
            \endsplit \tag 3-18
$$
Noting that $\pmb{ \psi}$'s are grassmannian variables,
the jacobian is given as
$(\delta \psi \delta \bar{\psi})/(\delta \psi' 
\delta \bar{\psi}')$.
In order to compute these variations (3-18), 
I will introduce complete sets associated with this system 
[29,30,34];
$$
  \ii\DD \varphi_n = \lambda_n \varphi_n , \ \ 
        (\rho \chi_n^\dagger) \ii \laDD = 
        \lambda_n \rho \chi_n^\dagger , \tag 3-19
$$
and
$$
        \int \rho \dd^2 q\  \chi_m^\dagger(q) \varphi_n (q)= 
        \delta_{m,n}. \tag 3-20
$$
Then the variation of the field is expressed as
$$
  {\pmb{ \psi}}'=:\sum_m a_m'\varphi_m = 
\sum_m e^{ -\alpha  }    a_m \varphi_m . \tag 3-21
$$

Here I will evaluate the fermionic jacobian in the 
transformations,
$$
        \split
        a_m' &= \sum_n \int \rho\dd^2 q \ 
        \chi_m^\dagger e^{-\alpha }
            \varphi_n a_n \\
            &=:\sum_n \Cal C_{m,n} a_n .
            \endsplit \tag 3-22
$$
The change of the functional measure is expressed by 
[29,30,32,33],
$$
    \prod_m \dd a_m' = [\roman{det}(\Cal C_{m,n})]^{-1}
                                \prod_m \dd a_m . \tag 3-23
$$
By calculation, the jacobian is written by more explicit form,
$$
    \split 
    [\roman{det}(\Cal C_{m,n})]^{-1}
      &=[\roman{det}(\delta_{m,n}-\int \rho\dd ^2q\ 
        \alpha\chi^\dagger_m (q) \varphi_n(q))]^{-1}\\
      &=\roman{exp}[ \sum_m \int\rho \dd^2 q\ 
        \alpha\chi^\dagger_m (q) \varphi_m(q)]\\
      &=:\roman{exp}[ \int \rho \dd^2 q\ 
        \alpha \Cal A(q) ]
     .   \endsplit \tag 3-24
$$
Since $\Cal A(q)$ is not well-defined and unphysically diverges,
  I must regularize it.
In this article, I will employ the modified negative power kernel
 regularization procedure
which is partially proposed by Alves {\it et al.} [35] and is a 
local version of 
the  Hurwitz $\zeta$ regularization [8,36,39].
However the Dirac operator $\DD$ is not hermite and the real 
part of some of the eigenvalues of $ -\DD^2$ are negative. 
Hence I cannot directly apply negative power regularization of 
 Alves {\it et al.} [35].
Even though the heat kernel function can be adapted for such 
Dirac operator with negative eigenvalues [39], 
$A(q)$ is not completely regularized by the heat kernel
as Alves {\it et al} pointed out [35]. 
Thus by  generalizing Hurwitz $\zeta$ function [36]
rather than the Riemann $\zeta$ function, I will modify 
the negative power regularization [35].

I will introduce a finite positive parameter
$$
      \mu^2> -\min_n (Re \lambda^2_n) \ge 0 , \tag 3-25
$$
and let the modified negative power kernel and the modified 
heat kernel  defined as [8,35,38,41], 
$$
     \Cal   K_\zeta(q,r,s|\mu)
        =\sum_m (\lambda_m^2+\mu^2)^{-s} 
        \varphi_m(q)\chi_m^\dagger(r)  ,\quad
 \Cal K_\HK(q,r,\tau|\mu)=\sum_m \ee^{-(\lambda_m^2+\mu^2)\tau}
                 \varphi_m(q)\chi_m^\dagger(r)     .
            \tag 3-26
$$
The both are connected by the Mellin transformation [35],
$$
      \Cal K_\zeta(q,q,s|\mu)=\frac{1}{\Gamma(s)}
      \int^\infty_0 \dd \tau \tau^{s-1}
      \Cal K_\HK(q,q,\tau|\mu). \tag 3-27
$$
From the definition, all quantities $\lambda_m^2+\mu^2$ 
are positive, the integration in (3-27) is well-defined.
If I also trace (integrate) $\Cal K_\zeta(q,q,s|\mu)$ 
over the space-time $q$, it is just a generalized
$\zeta$-function, which is generalization of Hurwitz 
$\zeta$ function for the Dirac operator,
$$
      \zeta(s,\mu)=\sum_m \frac{1}{(\lambda_m^2 +\mu^2)^s }.
       \tag 3-28
$$
Then $\Cal A(q)$ should be redefined,
$$
        \Cal A(q)\equiv \lim_{s \to 0} 
         \lim_{r \to q} \roman{tr}
          \Cal K_\zeta(q,r,s|\mu)
      .  \tag 3-29
$$

For small $\tau$, the heat kernel $K_\HK$ is
asymptotically expanded as [8,38],
$$
  \Cal  K_\HK(q,r,\tau|\mu)\sim\frac{1}{4\pi \tau}
            \roman{e}^{-(q-r)^2/4\tau}
            \sum_{n=0}^{\infty} e_n(q,r)\tau^n .
                                                   \tag 3-30
$$
Accordingly I calculate $\Cal  K_\zeta(q,q,s|\mu)$ as [35],
$$
      \split
      K_\zeta(q,q,s|\mu)&=\frac{1}{\Gamma(s)}\left(
      \int^\epsilon_0 \dd \tau \tau^{s-1}
      \Cal K_\HK(q,q,\tau|\mu)+
      \int^\infty_\epsilon \dd \tau \tau^{s-1}
      \Cal K_\HK(q,q,\tau|\mu)\right)\\
      &=\frac{1}{\Gamma(s)}\left(
      \int^\epsilon_0 \dd \tau \tau^{s-1}
      (\frac{1}{4 \pi \tau} \sum_n e_n \tau^n+
      \int^\infty_\epsilon \dd \tau \tau^{s-1}
      \Cal K_\HK(q,q,\tau|\mu)\right)\\
      &=\frac{1}{\Gamma(s+1)}\left(\frac{1}{4 \pi \tau}
       \sum_n e_n \frac{\epsilon^{n-1}}{s-1+n}
      + s G(s)\right). \endsplit\tag 3-31
$$
Here I used $\Gamma(s+1)=s \Gamma(s)$. 
Since $K_\HK(q,q,\tau|\mu) \propto \exp(- \lambda \tau) $
as $\tau \to \infty$ ($\lambda >0)$, the second term is a 
certain entire analytic function over the $s$-plane
and I denoted it $G(s)$. Noting $\Gamma(1)=1$, (3-29) turns out
$$
      \Cal A(q) = \frac{1}{4\pi} e_1 . \tag 3-32
$$
On the other hand, according to ref. [39], 
since the square of the Dirac operator (3-12) is given as
$$
      -  \DD^2 = \rho^{-1} \pmatrix -4\bar \partial \partial+
      2\rho^{-1}(\partial \rho) \bar \partial+ (K\rho-4p^2)
       &-  4 \rho^{1/2}(\bar \partial p\rho^{-1/2})\\
             - 4\rho^{1/2}( \partial p\rho^{-1/2})&
              -4\partial\bar\partial+
      2\rho^{-1}(\bar \partial \rho) \partial
                         +  (K\rho-4p^2)
                       \endpmatrix, \tag 3-33
$$  
the coefficient of the expansion (3-30) is written by,
$$
    \split
      e_1&= 4p^2\rho^{-1}- \mu^2+
      2\rho^{1/2}\sigma^a \delta_a^{\ \beta} \partial_\beta 
          p\rho^{-1/2}-\frac{5}{6}K\\
          &=  4p^2\rho^{-1} - \mu^2-\frac{5}{6}K
        + 2\rho^{1/2}\sigma^a \delta_a^{\ \beta} \partial_\beta 
          p\rho^{-1/2}. \endsplit
                \tag 3-34
$$
Noting the fact that trace over the spin index generates 
the functor $2$, I obtain,
$$
        \Cal A(q)=\frac{1}{2\pi} \left(4 p^2 \rho^{-1} 
        - \mu^2- \frac{5}{6} K\right)=\frac{1}{2 \pi} 
        \left(\frac{10}{3}\frac{1}{\rho} \partial 
        \bar \partial \phi- \mu^2
                  + H^2 \right),
   \tag 3-35
$$
and the jacobian,
$$
    \frac{\delta \psi \delta \bar{\psi}}
    {\delta \psi' \delta \bar{\psi}'}
            =\roman{exp}[  \int \rho \dd^2 q\ 
                        \alpha(q) \Cal A (q)] .  \tag 3-36
$$
I will derive the boson-fermion correspondence.
From (3-18), the Ward-Takahashi identity [29,30], 
$$
        \left. \frac{\delta}{\delta \alpha(q)}
        (Z_1-Z_2)\right|_{\alpha(q) = 0}\equiv0,      \tag 3-37
$$ 
gives an anomaly,
$$
    \rho^{-1} \delta_a^\alpha \partial_\alpha
    < \ii \bar{\pmb{ \psi}}\sigma^a \rho^{1/2}{\pmb{ \psi}}>
          +<\bar{\pmb{ \psi}}\ii \DD{\pmb{ \psi}}> 
        =  \frac{1}{2\pi}\mu^2+\frac{5}{12 \pi}
         K -\frac{1}{2\pi} (H^2)  , \tag 3-38
$$
where $<\Cal O>$ means the expectation value of $\Cal O$ 
related to the partition function (3-1).
I will refer this anomaly "immersed anomaly".

\tvskip
\centerline{\twobf \S 4. Discussion}
\tvskip

The right hand side of (3-38) is closely related to the 
conformal anomaly in the string theory
and the Liouville action.  If $H$ vanishes, 
the arguments in the previous section can be parallelled 
to the calculation of the conformal anomaly [22]. 
The case $H=0$ is known as the minimal surface in
the immersion geometry [2,11,27].
Thus the quantity $\mu^2$ introduced in (3-31) is identified 
with the coupling constant of 
the dilaton in the Liouville action [22].
This picture preserves in the region with the finite 
constant curvature $H$
and then the physical meaning of $\mu^2$ is clarified. 

Furthermore it should be noted that 
if I employ the heat kernel regularization instead of 
the modified negative power regularization, 
$\mu^2$ appears as infinite value, $\mu^2 \sim 1/\tau$.
Thus mathematically $\mu^2$ is interpreted as a 
convergence parameter which makes the 
kernel finite and this picture consists with the motivation 
to make the integral in (3-28) well-defined.

Here I will investigate the meanings of the anomalous relation
 (3-28) as follows.
I will integrate both sides in (3-39),
$$
      \int_\Sigma \dd^2 q 
      \left(\ii \delta_a^\alpha \partial_\alpha
      < \bar{\pmb{ \psi}}\sigma^a \rho^{1/2}{\pmb{ \psi}}>
          +\rho <\ii \bar{\pmb{ \psi}}\DD{\pmb{ \psi}}> 
       + \frac{1}{2\pi} \rho (H^2)
       -\frac{1}{2\pi}\mu^2\rho-\frac{5}{12 \pi}
        \rho K \right)=0. \tag 4-1
$$
The first term is locally expressed as total derivative
 $j:=<\ii \delta_a^\alpha \bar{\pmb{ \psi}}\sigma^a 
 \rho^{1/2}{\pmb{ \psi}}>\dd q^\alpha$. Thus let the surface
 $\Sigma$ be divided as
$$
      \Sigma = \Sigma_+ \cup \Sigma_-, \quad S^1 \approx 
      \Sigma_+ \cap \Sigma_- ,\quad
      \Sigma_+ \approx \Sigma_- \approx \Bbb R^2, \tag 4-2
$$
where $\approx$ means the homeomorphism and I will 
define $j_\pm$ as functions over $\Sigma_\pm$.
Then the integration of the first term becomes, 
$$
      \int_\Sigma \dd * j =\int_{\partial\Sigma_+ } ]
      *j_++\int_{\partial\Sigma_- }*j_-
      = \int_{\partial \Sigma_+ } (*j_+-*j_-)=B_2 \nu, \tag 4-3
$$
where $\nu$ is an integer and $B_2$ is a constant parameter.
Thus it can be regarded as the candidacy of the generator of 
the fundamental group of $\Sigma_+ \cap \Sigma_-$
while the Euler characteristic $\chi$ expresses the global
 topology of the surface.
If the current is conserved, $\nu$ vanishes.

Furthermore the third term means the area of the surface $\SS$,
$$
      A := \int_\Sigma \rho\  \dd^2 q. \tag 4-4
$$
Using these quantities, I obtain the global expression of (3-38),
$$
      < S_{\Dirac}>=\frac{1}{2\pi} 
      (\mu^2 A-W)+\frac{5}{6} \chi-B_2 \nu. \tag 4-5
$$
Even though the current is not conserved, $B_2 \nu$ is 
expected as a topological quantity.
Thus (4-5) means that the expectation value of 
the action of the Dirac operator is written as 
the Willmore functional and the area of the 
surface. If the mean curvature vanishes, 
the minimal of area of the surface corresponds
to stationary point of the action of the Dirac operator. 
This correspondence 
is theorem of the minimal surface and of old 
Weierstrass-Enneper equation [11].
For general immersion, investigation on the Dirac 
operator of the KKWE equation
(3-9) means studying the Willmore surface itself 
if fixing the area.
On the case of Schr\"odinger particle, the immersion 
effect appears as 
attractive potential and thus the sign of the Willmore 
action can be naturally
interpreted [43]. Furthermore,
since the Liouville action can be extended to that with 
supersymmetry [22],
I believe that this correspondence (4-5) between 
these actions  should be interpreted 
by supersymmetry of this system.

The Willmore surface problem of $\Bbb R^3$ has very similar 
structure of the elastica problem
of $\Bbb R^2$.
Corresponding to the Willmore functional (2-20), 
there is Euler-Bernoulli 
functional for an elastica [10,21],
$$
      E=\int \dd q^1\ k^2, \tag 4-6
$$
where $k$ is a curvature of the elastica [10,19]. While the 
Willmore surface is related to 
the modified Novikov-Veselov (MNV) equation, the elastica 
is related to the modified KdV equation.
From the soliton theory, the MNV equation is a higher 
dimensional analogue of the MKdV equation [6].
The Dirac operator appearing in the auxiliary linear 
problems of the MKdV equation is realized as the 
operator for the Dirac field confined in the elastica [28-32] as
the KKWE equation might be  related to auxiliary linear problems 
of the MNV equation
[2,6] and is realized as the equation of the Dirac field 
confined in the immersed surface [1].

In the series of works [28-32], I have been studying the 
elastica in terms of the quantized Dirac fields in
the elastica.
In terms of the partition function of the Dirac field, 
I constructed the Jimbo-Miwa theory
of the MKdV hierarchy [32] and showed the physical meaning 
of the inverse scattering method
and the monodromy preserving deformation [28,31].
Investigation on the Dirac operator of the KKWE equation 
might lead us to the Sato-type theory of the MNV hierarchy.

Furthermore, recently I exactly quantized the elastica of the 
Euler-Bernoulli functional (4-6) preserving its local length
and found that its moduli is closely related to the 
two-dimensional quantum gravity;
the quantized elastica obeys the MKdV hierarchy and at a critical 
point, the Painlev\'e equation of the first kind appears [21].
Instead of the local length preserving, after imposing that 
the surface 
preserves its complex structure or other constraints {\it e.g.} 
$\rho H=$constant [24,25], 
one could quantize the Willmore functional
and then, I expect that the MNV hierarchy might appear [44] as
the quantized motion of a Willmore surface in the path integral
 as the MKdV hierarchy appears in 
the quantization of the  elastica [21].

Moreover recently another relation between the geometry and
 quantum equation,
was discovered by Konopelchenko [45-47]. At this stage, 
I could not physically
interpret the new relation but believe that there is 
another quantum meanings.
I expect that his old and new relations [2-4,44-46] 
are clarified in the 
quantum mechanical context.

\tvskip
\centerline{\twobf Acknowledgment}
\tvskip

I would like to  thank Prof. S. Saito, for critical 
discussions and continuous encouragement. 
I am grateful to  Prof. Y. Ohnita for telling me the ref.[5] 
and ref.[44]
and to Prof. B.~G.~Konopelchenko for sending me his very 
interesting works and encouragement.
I would also like to thank  Prof.~T.~Tokihiro, Prof. 
K.~Sogo, Y. \^Onishi
and Prof. K.~Tamano
for helpful discussions at the earliest stage and 
continuous encouragement.


\Refs

\ref \no 1 \by S.~Matsutani \jour  J.  Phys. A: Math. \& Gen. 
\vol 30 \yr 1997 \pages4019-4029 \endref
\ref \no 2 \by B.~G.~Konopelchenko  \jour Studies in Appl.~Math.   
\vol 96  \yr1996 \pages 9-51 \endref
\ref \no 3 \by B.~G.~Konopelchenko and I.~A.~Taimanov 
\jour J.~Phys.~A: Math.~\& Gen.  
\vol 29  \yr1996 \pages 1261-65 \endref
\ref \no 4 \by B.~G.~Konopelchenko and I.~A.~Taimanov  \paper 
Generalized Weierstarass formulae, soliton equations 
and Willmore surfaces I.
Tori of revolution and the mKDV equation \jour dg-ga/9506011 
\endref

\ref \no 5 \by K.~Kenmotsu \jour Math. Ann. \year 1979 
\pages89-99 \vol 245 \endref

\ref \no 6 \by  I.~A.~Taimanov  
\paper Modified Novikov-Veselov equation and differential 
geometry of surface \jour dg-ga/9511005  
 \endref
\ref \no 7 \by I.~A.~Taimanov   
\paper Surface revolution in terms of soliton \jour 
dg-ga/9610013 \endref
\ref \no 8 \by N.~Berline, E.~Getzler and M.~Vergne, 
\book Heat Kernels and Dirac Operators
\publ Springer \yr 1991 \publaddr Berlin \endref
\ref \no 9 \by G.~Faltings \book Lectures on the Arithemtic 
Riemann-Roch theorem 
\publ Princeton Univ. Press \publaddr Princeton \yr 1992 \endref
\ref \no 10 \by C.~Truesdell \jour Bull. Amer. Math. Soc. 
\vol 9 \yr 1983 \pages 293-310 \endref
\ref \no 11 \by A.~I.~Bobenko \paper
Surfaces in terms of 2 by 2 matrices: Old and new integrable 
cases
\inbook Harmonic Maps and Integrable Systems 
\eds A.~P.~Fordy and J.~C.~Wood
\publ Vieweg \publaddr Wolfgang Nieger \yr 1994 \endref

\ref \no 12  \by R.~E.~Goldstein and D.~M.~Petrich
\jour Phys. Rev. Lett.\vol  67  \yr 1991, \pages3203-3206 \endref
\ref \no 13  \by R.~E.~Goldstein and D.~M.~Petrich
\jour Phys. Rev. Lett.\vol  67  \yr 1992, \pages555-558 \endref
\ref \no 14 \by S.~Matsutani \jour Int. J. Mod. Phys. A \vol 10 
\yr 1995 \pages3109-3130 \endref
\ref \no 15 \by A.~Doliwa, P.~M.~Santini \jour Phys. Lett. 
A \vol 185 \yr 1994 \pages 373-384 \endref
\ref \no 16 \by A.~Bobenko and U.~Pinkall, 
\jour J. Diff. Geom. \vol 43 \yr 1996 \pages 527-611\endref

\ref \no 17 \by H.~C.~Wente \jour Pacific J. Math   \vol 121 
 \yr1986 \page193-243\endref
\ref \no 18 \by U.~Abresh \jour J. reine u. angew Math.   
\vol 374  \yr1987 \page169-192\endref
\ref \no 19\by U.~Pinkall and I.~Sterling  \jour Ann. Math    
\vol 130  \yr1989 \page407-451\endref
\ref \no 20\by A.~P.~Fordy and J.~C.~Wood \book 
Harmonic Maps and Integrable Systems  
\publ Vieweg \publaddr Wolfgang Nieger \yr 1994 \endref
\ref \no 21 \by S.~Matsutani \jour solv-int/9707003 \endref

\ref \no 22 \by A.~M.~Polyakov  \book Gauge Fields and Strings
\publ Harwood Academic Publishers \yr 1987 \publaddr London 
\endref

\ref \no 23 \by J-L.~Gervais and Y.~Matsuo  
\jour Com. Math. Phys.  
\vol 152  \yr1993 \pages 317-368 \endref

\ref \no 24 \by R.~Carroll and B.~Konopelchenko 
\jour Int. J. Mod. Phys. \vol A11 \yr 1996 \pages1183-1216
\endref

\ref \no 25 \by K.~S.~Viswanathan and R.~Parthasarathy 
\jour Ann. Phys. \vol 244 \year 1995 \pages 241-261
\endref

\ref \no 26 \by T.~J.~Willmore \jour J. Lond. Math. Soc. \vol 2 
\yr 1971\pages307-310\endref

\ref \no 27 \by T.~J.~Willmore \book Riemannian Geometry 
\publ Oxford 
\yr 1993 \publaddr Oxford \endref

\ref \no 28 \by S.~Matsutani and H.~Tsuru \jour Phys. Rev A 
\vol 46 \yr1992\page1144-7 \endref

\ref \no 29 \by S.~Matsutani  \jour Prog. Theor. Phys. \vol 91 
\yr1994\page1005-37 \endref

\ref \no 30 \by S.~Matsutani \jour  J.  Phys. A: Math. \& Gen.
 \vol 28 \yr 1995 \pages1399-1412 \endref

\ref \no 31 \by S.~Matsutani 
\book Thesis in Tokyo Metropolitan Univ. 
\yr 1995  \endref

\ref \no 32 \by S.~Matsutani \jour Int. J. Mod. Phys. A \vol 10 
\yr 1995 \pages3109-3130 \endref

\ref \no 33 \by K.~Fujikawa\jour  Phys. Rev. Lett. 
\vol  42\year 1979\pages 1195-1199\endref

\ref \no 34\by  A. P. Balachandran, G. Marmo, V. P. Nair 
and C. G. Trahern \jour
Phys Rev D \vol  25\year 1982\pages 2713-2718\endref 

\ref \no 35 \by M.~S.~Alves, C.~Farina and C.~Wotzasek 
\jour Phys. Rev. D \vol 43 \year 1991 \pages4145-4147 \endref

\ref \no 36 \by E.~Elizalde, S.~D.~Odintsov, A.~Romeo,
A.~A.~Bytsenko and S.~Zerbini   \book 
Zeta Regularization Techniques with Apprications
\publ World Scientific \publaddr Singapore 
\yr 1994 \endref


\ref \no 37 \by M.~F.~Atiyah  and I.~M.~Singer  
\jour Ann. of Math.   
\vol 87 \yr1968 \page484-530 \endref

\ref \no 38 \by M.~F.~Atiyah  and I.~M.~Singer  
\jour Ann. of Math.  
 \vol 87 \yr1968 \page546-604 \endref

\ref \no 39 \by  P.~B.~Gilkey \book Invariance Theory, 
The Heat Equation and
the Atiyah-Singer Index Theorem
\publ Publish or Perish \yr 1984 \publaddr Wilmington \endref


\ref \no 40 \by A.~E.~H.~Love 
\book A Treatise on the Mathematical Theory of Elasticity
\publ Cambridge Univ. Press \yr 1927 \publaddr Cambridge \endref
\ref \no 41 \by P.~Ramond   \book Field Theory: A Modern Primer
\publ Benjamin \yr 1981 \publaddr Mento Park \endref

\ref \no 42 \by M.~Burgess and B.~Jensen   \jour Phys. Rev. A    
\vol48  \yr1993\page1861-6 \endref

\ref \no 43 \by S.~Matsutani \jour J.~Phys.~A:~Math.~\&~Gen. 
\vol 26 
\yr 1993 \page 5133-5143 \endref  


\ref \no 44 \by P.~G.~Grinevich and M.~U.~Schmidt \paper 
Conformal invariant functionals of immersioons of tori into 
$\Bbb R^3$ 
\jour dg-ga/9702015 \endref

\ref \no 45 \by B.~G.~Konopelchenko \jour Inverse Problem 
\vol 12 \yr 1996
\pages L13-L18 \endref


\ref \no 46 \by B.~G.~Konopelchenko \jour J. Math. Phys 
\vol 38 \yr 1997 \pages 434-543 \endref

\ref \no 47 \by R.~Beutler and B.~G.~Konopelchenko 
\paper Surfaces of Revolution via
the Schr\"odinger Equation: Construction, 
Integrable Dynamics and
Visualization \yr 1996 \endref

\endRefs
\enddocument